\documentstyle{article}
\title{Anisotropic Superconductivity of Quasi-One-Dimensional Electrons in Mag
netic Field}
\author{Mitake Miyazaki and Yasumasa Hasegawa\\
Faculty of Science,Himeji Institute of Technology, \\Kamigouri-cho, Akou-
gun, Hyogo 678-12, Japan }

\begin{document}
\sloppy
\maketitle

\begin{abstract}
We study theoretically the transition temperature $T_{\rm c}(H)$ of a qu
asi-one-dimensional anisotropic superconductor in a magnetic field, which is 
thought to be realized in organic conductors, (TMTSF$)_2$X. In the weak field 
limit, $-{\rm d}H_{\rm c2}/{\rm d}T$ for the anisotropic singlet is shown to 
be $\sqrt{3}$ times larger than that for the anisotropic triplet in the case 
of attractive interaction of electrons at nearest-sites along the $c$ axis 
and magnetic field along the $b$ axis. In the strong field region the 
transition temperature for spin-triplet is restored to zero field critical 
temperature while the an isotropic spin-singlet is suppressed by the Zeeman 
effect. There should be tra nsition from anisotropic singlet to anisotropic 
triplet as the magnetic field is increased.
\end{abstract}

\section{Introduction}

Recently, the reentrance of the superconductivity of quasi-one-dimensional 
electrons in a strong magnetic field attracts theoretical interest.
~\cite{rf:1,rf:2,rf:3,rf:4} The anomaly of the resistivity in a strong field is 
observed,~\cite{rf:5} which is thought to be a signal of the 
superconductivity. The superconductivity in a strong field is understood as 
follows. When the magnetic field is applied along the $b$ axis, which is 
different direction from the field-induced spin density wave (FISDW) case, 
the semi-classical orbits of electrons are localized in the $a$-$b$ plane 
(where $a$ and $b$ is 
the most conducting and second conducting axes, respectively), i.e. the 
motion along the $c$ axis is bounded. Since the orbital frustration due to 
the magnetic field comes from the motion of electrons in the plane 
perpendicular to the magnetic field, electrons can make Cooper pairs without 
affected by the orbital frustration in the strong field limit. The 
superconductivity in strong magnetic field can be also understood by nothing 
that the quasi-one-dimensional system in the magnetic field is essentially one
-dimension with the field-dependent effective interaction.~\cite{rf:6} 

Lebed'~\cite{rf:1} and Dupuis {\it et al}.~\cite{rf:2,rf:3,rf:4} have 
calculated the mean-field critical temperature of a quasi-one-dimensional 
superconductor. In these papers they considered both singlet and triplet 
pairings but they assumed the isotropic superconductivity, i.e. the energy 
gap is constant at the Fermi surface. However, considering the recent 
experimental results of NMR relaxation rate~\cite{rf:7}, impurity effect
~\cite{rf:8} and proximity to the SDW phase~\cite{rf:9} in the (TMTSF$)_2$X, 
the superconductivity in the quasi-one-dimensional organics is thought to be 
not the conventional type with the constant energy gap.
 
In the previous paper we have calculated the effective interactions in the 
magnetic field by assuming the attractive interaction between electrons at the 
nearest-sites in the $c$ direction, which results in the anisotropic 
superconductivity.~\cite{rf:6} For this case the transition temperatures for 
the spin-singlet and spin-triplet are degenerate in the absence of the 
magnetic field. We got the different magnetic field dependence of the 
superconductivity of anisotropic singlet and triplet. In that paper we took 
the quantum limit approximation (QLA), which is valid for the strong field. 
In the intermediate field, the transition temperature calculated in the QLA 
is known to be smaller than that without the QLA.~\cite{rf:10}

In this paper we study the transition temperature of the anisotropic 
superconductivity in a quasi-one-dimensional system by extending an elegant 
approach used by Lebed' and Dupuis {\it et al}. 

We also consider the effect of Pauli pair-breaking for an anisotropic 
superconductivity by taking account of the Zeeman term. In general, the 
critical field of spin-singlet superconductivity is decreased by the Pauli 
pair-breaking effect. When the number of quasi particle polarized by magnetic 
field exceeds a critical value, Cooper pairs (${\bf k}+{\bf q}/2 \uparrow, 
-{\bf k}+{\bf q}/2 \downarrow$) with total momentum ${\bf q}$ are 
stabilized rather than those of (${\bf k}\uparrow,-{\bf k}\downarrow$). The 
inhomogeneous superconductivity with spatial oscillation of the gap function 
is then possible, as was pointed out by Fulde and Ferrell~\cite{rf:11} and 
Larkin and Ovchinnikov~\cite{rf:12} (FFLO state). In three dimensional case 
with a spherical Fermi surface, the FFLO state is known to be possible only 
in the small region of magnetic field. However, in the one-dimensional case, 
the electron ($-{\bf k}+{\bf q}/2 \downarrow$) is always on the down-spins 
Fermi surface for any (${\bf k}+{\bf q}/2 \uparrow$) on the up-spin Fermi 
surface if we take an appropriate choice of {\bf q}.~\cite{rf:13,rf:14} As a 
result, at least a half of the Fermi surface is available for pairing, so 
that the reentrance in the strong field is not completely suppressed by the 
Pauli pair-breaking effect even for the spin-singlet pairing. 

\section{Gap equation}
 We consider the anisotropic superconductivity described by the dispersion 
 law (we take $\hbar=k_{\rm B}=1$ throughout the paper)
\begin{equation}
E_{k\sigma}=v_{\rm F}(|k_x|-k_{\rm F})-2t_b\cos(bk_y)-2t_c\cos(ck_z)+\sigma
\mu_{\rm B}H,\end{equation}
where $v_{\rm F}=2at_a\sin(ak_{\rm F})$ is the Fermi velocity for the motion 
along the $a$ axis, $t_a$, $t_b$ and $t_c$ are the transfer integrals along 
$a$, $b$ and $c$ axes, respectively, $a$, $b$ and $c$ are lattice constants 
and $\sigma\mu_{\rm B}H$ is the Zeeman energy for $\uparrow(\downarrow)$ spin 
($\sigma=+(-)$). For simplicity we take $a$, $b$ and $c$ axes to be 
perpendicular each other and to be along the $x$, $y$ and $z$ directions, 
respectively. 

For the non-interaction Hamiltonian ${\cal H}_0$ of a quasi-one-dimensional 
system we take account of the magnetic field by the Peierls substitution, 
i.e. ${\cal H}_0=E({\bf k}\rightarrow-i\nabla-e{\bf A})$. When the vector 
potential ${\bf A}$ is taken as ${\bf A}=(0,0,-Hx)$, the eigenstates and 
the corresponding eigenvalues are given by

\begin{equation}
\phi^{\alpha}_{\mbox{\scriptsize{\bf k}},\sigma}(x,l,m)={\rm e}^{{\rm i}k_xx+
{\rm i}k_ybl+{\rm i}k_zcm-{\rm i}\alpha\frac{2t_c}{\omega_{\rm c}}\sin(k_zc-
Gx)}
\end{equation}
and
\begin{equation}
\epsilon^{\alpha}_{\mbox{\scriptsize{\bf k}},\sigma}=v_{\rm F}(\alpha k_x-k_
{\rm F})-2t_b\cos(k_yb)+\sigma\mu_{\rm B}H,
\end{equation}
where the integers $l$ and $m$ label the cites in the directions $y$ and $z$, 
$\alpha=\mbox{sgn}(k_x)$ refers to the right/left sheet of the Fermi surface, 
and $\omega_{\rm c}=ev_{\rm F}cH$. The one-particle Green's function of a 
quasi-one-dimensional system is represented by
\begin{eqnarray}
G^{\alpha}_{\sigma}(x,x',k_y,k_z,\omega_n)&=&{\rm e}^{-{\rm i}\alpha\frac{2t_c}
{\omega_{\rm c}}[\sin(k_zc-Gx)-\sin(k_zc-Gx')]}\nonumber\\
& &\mbox{}\times \sum_{k_x}{\rm e}^{{\rm i}k_x
(x-x')}\tilde{G}^{\alpha}_{\sigma}(k_x,k_y,\omega_n)
\end{eqnarray}
with

\begin{equation}
\tilde{G}^{\alpha}_{\sigma}(k_x,k_y,\omega_n)=\frac{1}{{\rm i}\omega_n-v_{\rm 
F}(\alpha k_x -k_{\rm F})+2t_b\cos(k_yb)-\sigma \mu_{\rm B}H}
\end{equation}
where $\omega_n=(2n+1)\pi T$ is a Matsubara frequency. 

The interaction Hamiltonian is written as

\begin{equation}
{\cal H}_{\rm int}= \frac{1}{2}\sum_{\alpha,\alpha',\sigma,\sigma'}\sum_
{<i,j>}U_{ij}\psi^{\alpha'\dag}_{\sigma}({\bf r}_i)\psi^{\bar{\alpha}'\dag}_
{\sigma'}({\bf r}_j)\psi^{\bar{\alpha}}_{\sigma'}({\bf r}_j)\psi^{\alpha}_
{\sigma}({\bf r}_i),
\end{equation}
where $U_{ij}$ is the interaction between electrons at $i$ and $j$ sites and 
$\bar{\alpha}=-\alpha$. The $\psi^{\alpha}_{\sigma}({\bf r})$'s are 
operators for electrons moving on the sheet $\alpha$ of the Fermi surface:
\begin{equation}
\psi^{\alpha}_{\sigma}({\bf r})=\sum_{\alpha k_x>0}\sum_{k_y,k_z}
\phi^{\alpha}_{\mbox{\scriptsize{\bf k}},\sigma}({\bf r})C^{\alpha}_
{\mbox{\scriptsize{\bf k}},\sigma},
\end{equation} 
where $C^{\alpha}_{\mbox{\scriptsize{\bf k}},\sigma}$ ($C^{\alpha\dag}_{\mbox
{\scriptsize{\bf k}},\sigma}$) is the annihilation (creation) operator of an 
electron in the state $\phi^{\alpha}_{\mbox{\scriptsize{\bf k}},\sigma}
({\bf r})$.

In this paper we take the nearest-site interaction along $c$ axis and on-site 
interaction as
\begin{eqnarray}
U_{ij}=\left\{\begin{array}{ll}
U&\mbox{if}\ {\bf r}_i={\bf r}_j\pm c\hat{{\bf z}}\\
U_0&\mbox{if}\ {\bf r}_i={\bf r}_j\\
0&\mbox{otherwise}\hspace{2cm}.
\end{array}
\right.
\end{eqnarray}

We define the order parameter corresponding to spin-singlet pairing state 
\begin{equation}
\Delta^{({\rm s})}_{\uparrow\downarrow}({\bf r}_i,{\bf r}_j)=U_{ij}\sum_
{\alpha}\langle\psi^{\bar{\alpha}}_{\uparrow}({\bf r}_j)\psi^{\alpha}_
{\downarrow}({\bf r}_i)\rangle\nonumber
\end{equation}
and spin-triplet pairing state
\begin{equation}
\Delta^{({\rm t})}_{\sigma\sigma'}({\bf r}_i,{\bf r}_j)=U_{ij}\sum_{\alpha}
\langle\psi^{\bar{\alpha}}_{\sigma'}({\bf r}_j)\psi^{\alpha}_{\sigma}(
{\bf r}_i)\rangle\alpha . \nonumber
\end{equation}

In the mean field approximation, the linearized gap equation is obtained as 
\begin{equation}
\Delta(x,q_z)=\int_{|x-x'|>d} {\rm d}x'K(x,x',q_z)\Delta(x',q_z),
\end{equation}
where $d^{-1}$ is related to cutoff energy $\Omega$.
The kernel $K$ for the anisotropic spin-singlet ($\Delta=\Delta_{\uparrow
\downarrow}$) is given by
\begin{eqnarray}
K^{({\rm s})}(x,x',q_z)&=&-2UT\sum_{\omega_n}\sum_{\alpha,k_y,k_z}\cos^2(ck_z)
G^{\alpha}_{\uparrow}(x,x',k_y,k_z+\frac{q_z}{2},\omega_n)\nonumber\\
& &\mbox{}\times G^{\bar{\alpha}}_{\downarrow}(x,x',-k_y,-k_z+\frac{q_z}{2},
-\omega_n)\nonumber\\
&=&\frac{-UT}{bcv^2_{\rm F}}\frac{\cos[2\mu_{\rm B}H(x-x')/v_{\rm F}]}
{\sinh[|x-x'|2\pi T/v_{\rm F}]}\nonumber\\
& &\mbox{}\times\left[J_0\left(A(x,x',q_z)\right)+J_2\left
(A(x,x',q_z)\right)\right],
\end{eqnarray}
where $J_n(A)$ is the $n$th-order Bessel function, $G=eHc$, and
\begin{equation}
A(x,x',q_z)=\frac{8t_c}{\omega_{\rm c}}\sin[\frac{G}{2}(x-x')]\sin[q_z\frac{c}
{2}-\frac{G}{2}(x+x')].
\end{equation}

For the anisotropic spin-triplet ($\Delta=\Delta_{\uparrow\uparrow}$ or 
$\Delta_{\downarrow\downarrow}$) we get the kernel as
\begin{eqnarray}
K^{({\rm t})}(x,x',q_z)&=&-2UT\sum_{\omega_n}\sum_{\alpha,k_y,k_z}\sin^2(ck_z)
G^{\alpha}_{\sigma}(x,x',k_y,k_z+\frac{q_z}{2},\omega_n)\nonumber\\
& &\mbox{}\times G^{\bar{\alpha}}_{\sigma}(x,x',-k_y,-k_z+\frac{q_z}{2},
-\omega_n)\nonumber\\
&=&\frac{-UT}{bcv^2_{\rm F}}\frac{1}{\sinh[|x-x'|2\pi T/v_{\rm F}]}\nonumber\\
& &\mbox{}\times\left[J_0
\left(A(x,x',q_z)\right)-J_2\left(A(x,x',q_z)\right)\right],
\end{eqnarray}
while for the isotropic spin-singlet the kernel is given by
~\cite{rf:1,rf:2,rf:3}
\begin{equation}
K^{({\rm iso})}(x,x',q_z)=\frac{-U_0T}{bcv^2_{\rm F}}\frac{\cos[2\mu_{\rm B}
H(x-x')/v_{\rm F}]}{\sinh[|x-x'|2\pi T/v_{\rm F}]}J_0\left(A(x,x',q_z)\right).
\end{equation}
If the on-site interaction is repulsive $(U_0>0)$, the isotropic spin-singlet 
is not realized.

In the present assumption for the interaction the isotropic spin-triplet does 
not occur. The isotropic spin-triplet is possible only when the backward and 
forward scattering constants are different, which will be the case when the 
attractive interaction exists between electrons on the nearest-sites along 
the $a$ axis.

\subsection{Transition temperature for a weak field}
In the weak field limit ($\omega_{\rm c}\ll T$), where the magnetic length 
$2\pi/G$ is much larger than the thermal length $v_{\rm F}/2\pi T$, we can 
expand the order parameter and Bessel function in $(x-x')$. We can also 
neglect the Zeeman term in this limit. Then the second-order differential 
equation is given by
\begin{equation}
-v^2_{\rm F}\frac{\partial ^2\Delta}{\partial x^2}+4\eta t^2_c[1-\cos
(q_zc-2Gx)]\Delta=\frac{16\pi ^2}{7\zeta (3)}T^2_{\rm c0}\left(1-\frac
{T_{\rm c}}{T_{\rm c0}}\right)\Delta,
\end{equation}
where $T_{\rm c0}=(v_{\rm F}/\pi d)\exp(-\pi bcv_{\rm F}/U)$ is the 
transition temperature in the absence of magnetic field, $\zeta (z)$ is the 
Riemann's zeta function and
\begin{eqnarray}
\eta =\left\{\begin{array}{ll}
1&\mbox{for isotropic superconductivity}\\
\frac{1}{2}&\mbox{for anisotropic spin-singlet}\\
\frac{3}{2}&\mbox{for anisotropic spin-triplet}.
\end{array}
\right.
\end{eqnarray}
 When $T_{\rm c0}\ll t_c$, it can be further simplified by replacing the 
 periodic potential in the lhs by a set of decoupled harmonic potential 
 located at points ($q_zc/2+n\pi)/G$. Then we get the critical temperature as
\begin{equation}
T_{\rm c}(H)=T_{\rm c0}-\sqrt{2\eta}\frac{7\zeta (3)}{8\pi^2}\frac{t_cv_{\rm
 F}ec}{T_{\rm c0}}H,
\end{equation}
or
\begin{equation}
-\frac{{\rm d}H^b_{\rm c2}}{{\rm d}T}=\frac{1}{\sqrt{2\eta}}\frac{8\pi^2T
_{\rm c0}}{7\zeta (3)t_cv_{\rm F}ec}.
\end{equation}
The upper critical field $H^b_{\rm c2}$ in this region is $\sqrt{3}$ times 
larger for anisotropic spin-singlet than for anisotropic spin-triplet.
If the magnetic field is applied along the $c$ axis,
$-{\rm d}H^c_{\rm c2}/{\rm d}T=8\pi^2T_{\rm c0}
/(7\sqrt{2}\zeta (3)t_bv_{\rm F}eb)$, for both anisotropic spin-singlet and
spin-triplet in the weak field limit. Its value is the same as that for
the isotropic supeconductivity.
In contrast with this case, if the attractive interaction works between
electrons along $b$ axis,
 $-{\rm d}H_{\rm c2}^c/{\rm d}T=8\pi^2T_{\rm c0}/(7\sqrt{2\eta}\zeta3)t_bv_
 {\rm F}eb)$
 for the weak magnetic field along $c$ axis and
$-{\rm d}H_{\rm c2}^b/{\rm d}T=8\pi^2T_{\rm c0}/(7\sqrt{2}\zeta (3)t_cv_
{\rm F}ec)$ for the field
along $b$ axis.

We compare above results with the upper critical field
($-{\rm d}H^{b}_{\rm c2}/{\rm d}T=1.84$ and $-{\rm d}H^c_{\rm c2}/{\rm d}
T=0.111$[T/K]) observed by
Murata {\it et al}~\cite{rf:15}
for (TMTSF$)_2$ClO$\mbox{}_4$. When we take the parameters as $b=7.7\AA$,
$c=13.5\AA$, the
 ratio of transfer integrals are $t_c/t_b=0.034$ if we fit the
experiments with the
 isotropic superconductivity. If we assume the attractive interaction
along the $c$ axis as studied in this paper we get $t_c/t_b=0.048$. In
the case of attractive
 interaction along the $b$ axis, we get $t_c/t_b=0.028$ to fit the
experiment.

\subsection{Transition temperature for any magnetic field}
Without using the weak field approximation studied in the previous section, 
we can solve the linearized gap equation eq.(11) numerically as done by 
Dupuis and Montambaux~\cite{rf:2,rf:3} for isotropic superconductivity. 
We write
\begin{equation}
\Delta_Q(x)={\rm e}^{{\rm i}Qx}\sum_l\Delta^{Q}_{2l}{\rm e}^{{\rm i}2lGx},
\end{equation}
where Bloch wave vector $Q$ is taken as $-G<Q\le G$. Then eq.(11) is written 
as a matrix equation

\begin{equation}
\Delta^Q_{2l}=\sum_{l'}A^Q_{2l,2l'}\Delta^Q_{2l'},
\end{equation}
where
\begin{equation}
A^Q_{2l,2l'}=\sum_NK_{N-2l,N-2l'}\tilde{K}(Q+NG),
\end{equation}
In the above $\tilde{K}(q_x)$ is given by
\begin{eqnarray}
\tilde{K}(q_x)&=&T\sum_{\omega_n}\sum_{\alpha,k_x,k_y}\tilde{G}^{\alpha}_
{\sigma}(k_x,k_y,\omega_n)\tilde{G}^{\bar{\alpha}}_{\bar{\sigma}}
(q_x-k_x,-k_y,-\omega_n)\nonumber\\
&=&\frac{cN(0)}{2}\sum_{\alpha}\left[\ln\left(\frac{2\Omega\gamma}{\pi T}
\right)+\Psi\left(\frac{1}{2}\right)\right.\nonumber\\
& &\left.\mbox{}-\mbox{Re}\Psi\left(\frac{1}{2}+\frac
{\alpha v_{\rm F}q_x+2\mu_{\rm B}H}{4{\rm i}\pi T}\right)\right],
\end{eqnarray}
for the spin-singlet, where $N(0)=1/\pi v_{\rm F}bc$ is the density of states 
for one spin at the Fermi energy, $\gamma$ is the exponential of the Euler 
constant, $\Omega$ is the cut-off energy, and $\Psi$ is the digamma function. 
For the spin-triplet, Pauli pair breaking term $2\mu_{\rm B}H$ in the digamma 
function should be eliminated.

The coefficients $K_{N_1,N_2}$ are given for the anisotropic singlet as

\begin{eqnarray}
K^{({\rm s})}_{N_1,N_2}&=&\frac{-2U}{c}{\rm e}^{{\rm i}\frac{\pi}{2}(N_1-N_2)}
\int^{2\pi}_0\frac{{\rm d}u}{2\pi}\cos^2(u)J_{N_1}\left(\frac{4t_c}{\omega_
{\rm c}}\sin(u)\right)\nonumber\\
& &\mbox{}\times J_{N_2}\left(\frac{4t_c}{\omega_{\rm c}}\sin(u)\right),
\end{eqnarray}
and for the anisotropic triplet as
\begin{eqnarray}
K^{({\rm t})}_{N_1,N_2}&=&\frac{-2U}{c}{\rm e}^{{\rm i}\frac{\pi}{2}(N_1-N_2)}
\int^{2\pi}_0\frac{{\rm d}u}{2\pi}\sin^2(u)J_{N_1}\left(\frac{4t_c}{\omega_
{\rm c}}\sin(u)\right)\nonumber\\
& &\mbox{}\times J_{N_2}\left(\frac{4t_c}{\omega_{\rm c}}\sin(u)\right),
\end{eqnarray}
while for the isotropic singlet
\begin{equation}
K^{({\rm iso})}_{N_1,N_2}=\frac{-U_0}{c}{\rm e}^{{\rm i}\frac{\pi}{2}
(N_1-N_2)}\int^{2\pi}_0\frac{{\rm d}u}{2\pi}J_{N_1}\left(\frac{4t_c}{\omega_
{\rm c}}\sin(u)\right)J_{N_2}\left(\frac{4t_c}{\omega_{\rm c}}\sin(u)\right).
\end{equation}
If we take the most divergent term ($v_{\rm F}(Q+NG)+2\mu_{\rm B}H=0$) 
and neglect the others in eq.(22), we get the result in QLA.

\subsubsection*{A. Without Zeeman effect}
We first consider the case without Pauli pair breaking. In Fig.1 the 
transition temperature for anisotropic and isotropic spin-singlet is shown 
as a function of $ h=H/H_0$, where $H_0=4t_c/v_{\rm F}ebc$.~\cite{rf:16} 
We take parameters for the organic conductor as $a\approx7.3\AA$, 
$c\approx13.5\AA$, $t_a\approx2900$K, $t_c\approx17$K and $ak_{\rm F}=\pi/4$, 
and get $H_0\approx11.4$T. Here two transition lines are characterized by 
the order parameter for $Q=0$ or $Q=G$. The very small oscillation of the 
transition line for the anisotropic spin-singlet exists in
$0.05<h<0.25$ in the present choice of parameters.
The recovering to zero field critical temperature starts from lower field 
in the anisotropic spin-singlet case than in the isotropic case. In Fig.2 we
plot the transition temperature for anisotropic and isotropic triplet
with the same parameters as in Fig.1. The magnetic field dependence of
transition temperature for isotropic spin-triplet is the same as that
for isotropic spin-singlet. In the case for the anisotropic spin-triplet, 
the oscillation is large and the recovering starts from higher field than 
the isotropic type (Fig.2). 

These results are consistent with those obtained in QLA.~\cite{rf:6}

\subsubsection*{B. Zeeman effect}
We next consider the effect of Pauli pair-breaking 
by taking account of the Zeeman term. As seen in eq.(23), logarithmic 
divergences occur at $Q=\pm2\mu_{\rm B}H/v_{\rm F}$ and $Q=
\pm(G-2\mu_{\rm B}H/v_{\rm F})$ even in the presence of the Zeeman term.  
We take these values of Q and calculate the transition temperature. As 
shown in Fig.3, the transition temperature of the spin-singlet is 
considerably reduced by the Pauli pair-breaking effect. The transition
temperature for $Q=\pm(G-2\mu_{\rm B}H/v_{\rm F})$ is too small to be 
seen for $h>0.1$ and almost
degenerate with that for $Q=\pm2\mu_{\rm B}H/v_{\rm F}$ for $h<0.1$ 
in Fig.3. In the case of equal-spin-pairing $\Delta_{\uparrow\uparrow}$ 
and $\Delta_{\downarrow\downarrow}$, Zeeman splitting does not change 
the transition temperature. Thus, one will observe spin-triplet 
superconductivity in the strong field.

Even with the Zeeman effect, Cooper pairs of whole electrons in the 
Fermi surface contribute to superconductivity if the N($\uparrow$)th energy 
level coincides with the N+1($\downarrow$)th level by Zeeman splitting. 
In this case the Pauli pair breaking effect becomes not important, as shown 
in Fig.4. This is the case when $2\mu_{\rm B}H = Gv_{\rm F}$  as seen in 
eq.(23), which yields $act_a=\phi_0\mu_{\rm B}/\sqrt{2}\pi$, where 
$\phi_0=hc_0/e$ is the flux quantum. This condition is satisfied if 
$act_a$ is about $1/10$ times of its value for organic conductors. This may 
be possible for a system with a large effective mass, or in the tilted 
magnetic field in the $a$-$b$ plane, because the Zeeman splitting is in 
proportion to the magnetic field, while magnetic length with respect to 
orbital motion is in proportion to perpendicular component of the 
magnetic field. On the other hand, in the case of tilted magnetic field in 
the $b$-$c$ plane, since hopping along $b$ axis is suppressed at the same 
time, one has to consider competition between superconductivity and FISDW.

In the weak field limit the field dependence of the critical temperature is 
given by the parameter $t_cv_{\rm F}\propto t_ct_a$, while in the strong 
field region it is given by $H_0\propto t_c/v_{\rm F}\propto t_c/t_a$ and 
$T_{\rm c0}$. In order to fit the observed initial slope $-{\rm d}H^b_
{\rm c2}/{\rm d}T |_{T_{\rm c0}}$~\cite{rf:3}, we take $t_at_c\sim 2100
\mbox{K}^2$ for the anisotropic singlet case and $t_at_c\sim 1500\mbox
{K}^2$ for the isotropic singlet case. In Fig.5 we plot the transition 
temperature for the isotropic singlet and isotropic triplet by taking 
$t_c=1.5$K and $T_{\rm c0}=1.36$K. If we assume that the transition 
temperature for isotropic triplet in the
absence of magnetic field is much smaller than that for the isotropic
singlet, the result seems to be consistent with experiment as discussed
by Dupuis~\cite{rf:4} and Lee {\it et al}.~\cite{rf:5} In Fig.6 we plot the 
critical temperature for the anisotropic singlet and anisotropic triplet as 
a function of the magnetic field by taking $t_c=1.5$K and $T_{\rm c0}=1.36$K. 
In this choice of parameters oscillation of the transition temperature for 
anisotropic triplet is pushed to the lower field. The transition temperature 
for the anisotropic spin-singlet with $Q=G-2\mu_{\rm B}H/v_{\rm F}$ changes
 discontinuously to very small value at $H=H_{\rm c}\sim 3.1$T. By changing 
 $Q$ we may get the continuous transition temperature in $H\geq H_{\rm c}$, 
 but this will be smaller than the transition temperature for $Q=2\mu_{\rm B}
 H/v_{\rm F}$. Thus we did not consider the change of $Q$. As shown in Fig.6, 
 the transition from anisotropic spin-singlet to anisotropic spin-triplet
occurs about $H=1.5$T and the transition temperature increases above
this field. This feature is not observed in the experiment.~\cite{rf:5}
This may be due to the assumption of the attractive interaction we have
used in this paper. The part of attractive interaction would be caused
by the anti-ferromagnetic spin fluctuation, which results in the spin-
dependent effective interaction on the nearest-sites and favors the 
anisotropic singlet pairing rather than anisotropic triplet pairing. Then the
transition temperature for the anisotropic triplet becomes lower and the
transition from anisotropic singlet to anisotropic triplet will occur at
higher magnetic field. Then we can also explain the experiment by Lee et
 al. with the anisotropic superconductivity.

\section{Conclusion}
In this paper we have studied the transition temperature for a quasi-one-
dimensional anisotropic superconductivity in a magnetic field applied along 
the $b$ axis. Here we have assumed the nearest-site attractive interaction 
between electrons along $c$ axis, which is independent of spin. The 
transition temperature for the anisotropic spin-singlet is same as that for 
the anisotropic spin-triplet
in the absence of magnetic field. We have shown that the critical field for 
the anisotropic spin-singlet is $\sqrt
{3}$ times larger than that for the anisotropic spin-triplet in the weak
magnetic region.

We have calculated the transition line of spin-singlet superconductivity 
considering the Zeeman effect. Although spin-singlet sate exists even with 
the Zeeman effect, the transition temperature is suppressed. On the other 
hand, spin-triplet superconductivity is not affected by the Zeeman effect. 
Therefore, the transition from spin-singlet to spin-triplet will occur as 
the magnetic field is increased. 

The present results for anisotropic spin-singlet are
consistent with the recent experiment by Lee {\it et al} ~\cite{rf:5}, which
shows the signal for high field superconductivity, if we take into
account 
the attractive interaction for the anisotropic spin-singlet and the
relative suppression of the anisotropic spin-triplet caused by the
anti-ferromagnetic spin fluctuation. Even in this case the transition
from anisotropic spin-singlet to anisotropic spin-triplet will occur in
the strong magnetic field as long as the spin-triplet is not destroyed
completely.

\section*{Acknowledgments}
This work is financially supported by the Grant-in-Aid for Scientific 
Research on the priority area ''Novel Electronic States in Molecular 
Conductor'' from the Ministry of Education.

\section*{Figures}
Fig.1. Critical temperature $T_{\rm c}(H)/T_{\rm c0}$ vs magnetic field 
along the $b$ axis for the anisotropic singlet (solid lines) and isotropic 
singlet (broken
lines). The Zeeman effect is neglected.

Fig.2. Critical temperature $T_{\rm c}(H)/T_{\rm c0}$ vs magnetic field for 
the
anisotropic triplet (solid lines) and the isotropic singlet (broken
lines).

Fig.3. The same as Fig.1 with the Zeeman effect.

Fig.4. The solid lines show Energy spectrum of a quasi-one-dimensional 
system in a magnetic field and the broken lines show the Zeeman splitting.

Fig.5. Critical temperature vs magnetic field for the isotropic singlet
(solid lines) and isotropic triplet (broken lines). Parameters are taken to 
fit the initial slope of the critical field $-{\rm d}H^b_{\rm c2}/{\rm d}T$ 
to the
experiment and different from that in Figs.1$\sim$3.

Fig.6. Critical temperature $T_{\rm c}(H)/T_{\rm c0}$ vs magnetic field 
for the
anisotropic spin-singlet (solid lines) and for the anisotropic
spin-triplet (broken lines) with the different choice of parameters
 from that in Figs.1$\sim$3. The Zeeman effect is taken into account.
\end{document}